\documentclass[11pt, twocolumn]{article}

\usepackage{geometry}
\usepackage{setspace}
	
\usepackage[utf8]{inputenc}
\usepackage[american]{babel}
\usepackage[T1]{fontenc}
\usepackage{microtype}

\usepackage{authblk}
\usepackage{nth}
\usepackage[inline]{enumitem}
\usepackage{amsmath}
\usepackage{amssymb}
\usepackage{commath}
\usepackage[exponent-product=\cdot,retain-unity-mantissa=false]{siunitx}
\usepackage{miller}

\usepackage{caption}
\usepackage{subcaption}

\usepackage{xr}
\usepackage[hyperref,fixpdftex]{xcolor}
\usepackage{graphicx}
\DeclareGraphicsExtensions{.pdf,.png}
\graphicspath{{figures/}}

\usepackage{caption}
\usepackage{csquotes}

\usepackage[%
	backend=biber,
	style=numeric-comp,
	giveninits=false,  
	sorting=none,
	doi=false,
	isbn=false,
	eprint=false,
	url=false,
]{biblatex}

\usepackage{xcolor}
\definecolor{shadecolor}{RGB}{220,220,220}

\usepackage{hyperref}
\hypersetup{
    hidelinks = true,
	colorlinks = false,
	linkcolor = {red!80!black},
	anchorcolor = {black},
	citecolor = {green!80!black},
	filecolor = {cyan!80!black},
	menucolor = {red!80!black},
	runcolor = {cyan!80!black},
	urlcolor = {magenta!80!black},
}
\usepackage[noabbrev]{cleveref}
\crefname{figure}{Figure}{Figures}
\crefname{table}{Table}{tables}

\usepackage[shortcuts=abbreviations]{glossaries-extra}
\setabbreviationstyle{long-short}  

\newabbreviation{transmission electron microscopy}{TEM}{transmission electron microscopy}
\newabbreviation{x-ray diffraction}{XRD}{x-ray diffraction}
\newabbreviation{focused ion beam}{FIB}{focused ion beam}
\newabbreviation{scanning transmission electron microscopy}{STEM}{scanning \glsxtrshort{transmission electron microscopy}}
\newabbreviation{identifier}{id}{identifier}
\newabbreviation{body-centered cubic}{BCC}{body-centered cubic}
\newabbreviation{face-centered cubic}{FCC}{face-centered cubic}
\newabbreviation{finite element method}{FEM}{finite element method}
\newabbreviation{discrete dislocation dynamics}{DDD}{discrete dislocation dynamics}
\newabbreviation{Peach-Koehler force}{PK-force}{Peach-Koehler force}

\usepackage{titlesec}
\titlespacing*{\section}{0pt}{0.9\baselineskip}{0\baselineskip}
\titlespacing*{\subsection}{0pt}{0.9\baselineskip}{0\baselineskip}
\titlespacing*{\subsubsection}{0pt}{0.5\baselineskip}{0\baselineskip}

\titleformat{\section}{\normalsize\bfseries\uppercase}{\thesection}{1em}{}
\titleformat{\subsection}{\normalsize\bfseries}{\thesubsection}{1em}{}
\titleformat{\subsubsection}{\normalsize\em}{\thesection}{0em}{}

\renewcommand\vec[1]{\ensuremath{\boldsymbol{#1}}}
\providecommand{\crystaldirection}[1]{\hkl[#1]}

\providecommand{\crystalplane}[1]{\hkl(#1)}

\usepackage{xspace} 
\newcommand\framename{Frame}
\newcommand\framefigref{\cref{fig:analyzed-frames}}
\newcommand\frameA{\framename\ \num{4727} (\framefigref a)\xspace}

\newcommand\frameC{\framename\ \num{4788} (\framefigref c)\xspace}

\newcommand\frameK{\framename\ \num{1}    (\framefigref k)\xspace}

\newcommand\frameN{\framename\ \num{302}  (\framefigref n)\xspace}
\newcommand\frameO{\framename\ \num{308}  (\framefigref o)\xspace}

\usepackage{ulem}
\definecolor{RussianGreen}{HTML}{729b79}
\definecolor{AirSuperiorityBlue}{HTML}{79aecd}
\definecolor{OceanboatBlue}{rgb}{0.0, 0.47, 0.75}  
\definecolor{Lilac}{HTML}{C3ACCE}  
\definecolor{RoseMadder}{HTML}{da2c38}  
\definecolor{RubyRed}{HTML}{931621}  
\definecolor{BrightMaroon}{HTML}{b5123e}  
\definecolor{AlloyOrange}{HTML}{c16200}
\definecolor{MaximumGreen}{HTML}{698f3f}
\definecolor{GenericViridian}{HTML}{2A7F62}
\definecolor{SlimyGreen}{HTML}{1A962B} 
\definecolor{LaSalleGreen}{rgb}{0.03, 0.47, 0.19}  

\newcommand\oldadd[1]{#1}
\newcommand\olddelete[1]{{}} 

\newcommand\add[1]{#1}
\newcommand\delete[1]{}
\newcommand\replace[2]{#2}

\usepackage[final]{changes}
\definechangesauthor[name=Sandfeld, color=LaSalleGreen]{StS}

\title{\bf\rmfamily
{{\bf \rmfamily Data-Mining of In-Situ TEM Experiments: Towards Understanding Nanoscale Fracture}}
}

\author[1]{Dominik Steinberger}
\author[2]{Inas Issa}
\author[1, 3]{Rachel Strobl}
\author[2, 4]{Peter J. Imrich}
\author[2, *]{Daniel Kiener}
\author[1, 3, 5, *]{Stefan Sandfeld}

\affil[1]{Institute of Mechanics and Fluid Dynamics, Freiberg University of Mining and Technology, Freiberg, Germany}
\affil[2]{Department Materials Science, Chair of Materials Physics, Montanuniversität Leoben, Leoben, Austria}
\affil[3]{Institute for Advanced Simulations -- Materials Data Science and Informatics (IAS-9), 
	      Forschungszentrum J\"ulich GmbH, 52425 J\"ulich, Germany}
\affil[4]{
Now at Kompetenzzentrum Automobil- und Industrieelektronik GmbH, Europastraße 8, 9524 Villach, Austria}

\affil[5]{Chair of Materials Data Science and Materials Informatics, Faculty 5 -- Georesources and Materials Engineering, 
	      RWTH Aachen University, 52056 Aachen, Germany}
\affil[*]{corresponding authors}

\date{}

\begin{document}
\sloppy
\twocolumn[
\begin{@twocolumnfalse}
	\renewcommand{\abstractname}{\vspace{-2.5\baselineskip}}
	\maketitle
	\begin{center}
	\begin{abstract}
			\begin{minipage}{0.87\textwidth}
	    The lifetime and performance of any engineering component, from nanoscale sensors to macroscopic structures, are strongly influenced by fracture processes. Fracture itself is a highly localized event; originating at the atomic scale by bond breaking between individual atoms close to the crack tip. These processes, however, interact with defects such as dislocations or grain boundaries and influence phenomena on much larger length scales, ultimately giving rise to macroscopic behavior and engineering-scale fracture properties. This complex interplay is the fundamental reason why identifying the atomistic structural and energetic processes occurring at a crack tip remains a longstanding and still unsolved challenge.
	    
	    We develop a \replaced{new analysis approach}{completely new approach} for combining quantitative \emph{in-situ} observations of nanoscale deformation processes at a crack tip with three-dimensional reconstruction of the dislocation structure and advanced computational analysis to address \added{plasticity and} fracture initiation in a ductile metal. Our combinatorial approach reveals details of dislocation nucleation, their interaction process, and the local internal stress state, all of which were previously inaccessible to experiments. This enables us to describe fracture processes based on local crack driving forces on a dislocation level \replaced{with a high fidelity that paves the way towards a better understanding and control of local failure processes in materials}{a previously unseen high fidelity that paves the way  to better understand and control local failure processes in materials}.
	    
	    \end{minipage}
	\end{abstract}
	\end{center}
	\vspace{1.5\baselineskip}
\end{@twocolumnfalse}
]

\section{Introduction}
\label{sec:introduction}

Plasticity is mainly carried by dislocations---line-like defects in the crystal lattice of materials with a ``core'' radius of  $\approx$\SI{1}{\nano\meter}. Though seemingly small, their collective behavior is responsible for a wide range of emergent properties: ultimately, dislocations determine the life time of turbine blades \autocite{Clemens2013_AEM15}, influence the performance of semi-conductors \autocite{Gudiksen2002_Nature415, Brune1998_Nature394}, or the degradation of energy storage materials \autocite{Singer2018_NatureEnergy3}.
Although metal plasticity has been technologically exploited for millennia, dislocations were not identified until the early 20th century. Their complex nature implies that, depending on scale and viewpoint of observation, it may be appropriate to envisage them as atomistic defects \olddelete{in the } in the crystal lattice, as localized plastic strain incompatibilities and line-like elastic singularities, as moving and interacting mathematical curves, or as mesoscale continuum fields. This inherently multi-faceted nature renders a universal description of the interaction among themselves or with the materials’ microstructure, as well as with other, e.g., geometrical details of devices, a highly demanding task---but crucial for the optimal design of new engineering materials. This is of particular importance during ductile fracture processes, where the interplay between individual dislocations and a highly localized crack tip becomes decisive for the reliability and lifetime of any device or structure.

Powerful microscopes allow us to resolve many important details of dislocations: high-resolution electron backscatter diffraction methods can reconstruct so-called geometrically necessary dislocation (GND) densities in averaging volumes of well below \SI{1}{\micro\meter\tothe{3}} \cite{Wallis2016_Ultramicroscopy168}, and today even resolve individual near surface dislocations when using electron channeling or transmission setups \autocite{gianola_britton_zaefferer_2019}.
%
X-ray diffraction mainly reveals total and GND densities and lattice strain with typical resolutions of well above \SI{1}{\micro\meter\tothe{3}} \autocites{Ribarik_2010_a}, unless nanofocused X-ray beams are employed in conjunction with very thin specimens at large scale research facilities. However, in their common implementations, both methods only extract incomplete and averaged information about dislocations and internal stresses, as opposed to \ab{transmission electron microscopy}, which conveniently reveals the position and nature of individual dislocations \autocites{Girault_2010_a,Kiener2011,Legros2014}. Furthermore, there have been significant advances in local strain analysis, extending the accessible region of interest from a few couple of nanometers based on high resolution images to large micrometer-sized fields of view by using 4D STEM approaches. However, for arbitrary crystal orientations, highly distorted crystal regions, or pronounced localized plasticity, local stresses can still only be roughly estimated (if at all) due to the experimental challenges of local strain measurement as well as the conceptual complication of a presumed stress-strain relation in a plastically deforming material \autocite{Gammer2016_APL109}. A further problem arising for any kind of analysis algorithm is the fact that dislocations in TEM images are essentially 2D projected ``pixel clusters'' of 3D objects. Due to the prevalent local diffraction conditions, they can exhibit pronounced intensity fluctuations along their path which complicates segmentation. Furthermore, due to their 3D nature, they cannot be simply treated as geometrical lines in 2D for which local stresses and energies could be inferred. These aspects render relating the microstructural details to underlying mechanisms and deformation behaviors on the specimen scale a challenging task.

However, the introduction and easy accessibility of powerful computational concepts and predictive models has initiated a paradigm shift in materials science: data-driven approaches can be used to guide experiments towards discovering unknown mechanisms, enabling the design of advanced materials \autocites{SAMAEE2020138295, Meiners2020_Nature2020, Hajilounezhad2021_CompMat7}, high-throughput screening helps to identify specific data sets \autocite{Chen2019_CompMater5}, machine learning and big data analyses predict materials’ properties with applications ranging from quantum chemistry to alloy design \autocites{Reuber_2014_a, mgi_2016, Banko2021}.
However, methods that bridge between experiments and simulations are still rare, and a detailed analysis and quantitative characterization of dislocations on multiple length scales from experiments and simulations lag behind those computational developments.

Turning to the fundamentals of plasticity and fracture processes, remarkable progress was achieved in recent years using miniaturized quantitative testing setups.
While these designs allow to conduct a quantitative experiment at length scale ranging from several tens to $\approx$\SI{100}{\nano\meter} \autocites{Haque2010,Minor2006,Kiener2011,Legros2014,Ma2010}, from a fracture mechanical viewpoint only the global loading parameters are known \autocites{Beaber2010,Gludovatz2014}.
This is further complicated by the fact that the multiaxial stress state at a crack tip is expected to activate multiple slip systems, rendering local TEM based analysis very demanding \autocites{Tanaka2008}.
From the computational viewpoint, atomistic simulations lend themselves to studying fracture processes, as the crack geometry can be easily designed with appropriate boundary conditions to address atomistic processes \autocites{Warner2007, Song2013, Moller2015, Bitzek2015}.
However, with respect to advanced modern high strength-high toughness materials \autocites{Ritchie2011,Gludovatz2014}, these modeling techniques still bear a number of shortcomings, as they operate at vastly different strain rates, are often not able to treat realistic volumes and alloy compositions, and therefore can be problematic in linking the nano scale observations to the specimen or device dimensions.\\

In the present work, dedicated quantitative \emph{in-situ} TEM fracture experiments were conducted that allow for detailed and highly controlled observation of crack tip processes, such as dislocation nucleation. \olddelete{
	These were closely interlinked with a new computational data mining framework, which -- based on digitized TEM images -- enabled to reconstruct the three-dimensional dislocation structure. This allows direct use of such image information as input for dislocation simulations that serve as a high fidelity, computer-based analysis of the experimental data. 
}\
\oldadd{
	These were closely interlinked with a new computational framework, which consists of (i) digitizing TEM images followed by (ii) reconstructing the three-dimensional dislocation structure which (iii) makes it possible to use such dislocation structures as input for dislocation simulations. (iv) Systematic sampling the parameter space of possible slip systems and dislocation characters then allows for data mining which in turn allows to extract information from the experiment in an unbiased, automated analysis of the experimental data. 
}

Our tightly coupled computational-experimental setup accounts for elastic anisotropy and  dislocation interactions with free surfaces. \add{It is therefore a representation of the experiment that is detailed in many relevant aspects and that, e.g., provides}\ access to local driving forces and energies acting on  the crack as well as on individual dislocations. Thereby, we lay the foundation for a better understanding of the complex interaction between crack geometry, dislocation nucleation and crystallographic details.

In the following, we start by highlighting the most important observations and results of the \emph{in-situ} \ab{transmission electron microscopy} fracture experiment. In a next step, the \ab{transmission electron microscopy} micrographs containing dislocation structures are digitized, the three-dimensional dislocation microstructure is reconstructed, 
\added{
	and detailed finite element analyses on the nanometer scale are performed for obtaining the dislocation-based stress state of the sample, as explained in section \enquote{Materials And Methods} below. 
	To resolve the ambiguity in the experiment concerning the exact slip system on which dislocations are nucleated and to determine their line sense, a data-mining approach is used through which probable or inpossible slip systems and line senses can be inferred. The implications of these results and the potential of such approaches are discussed with emphasis on future developments in the field.
}
\deleted{
	To address the inherent ambiguity concerning the slip system on which dislocations are nucleated, one of the main achievements of this work is the analysis of the three-dimensional dislocation structure as well as the exact identification of the experimental slip system and the dislocation line sense. The implications of these results are discussed with emphasis on future developments in the field. 
}

\section{Materials and Methods}
\label{sec:methods}
In the following, the sample preparation, the  experimental methods and the computational methods are introduced. 
Additional information can be found in the supplementary material.
%
%
\subsection{Experiment}
\label{sec:experiment}

\begin{figure}
	\centering
	\hspace{0.2em}
	\includegraphics[width=\columnwidth]{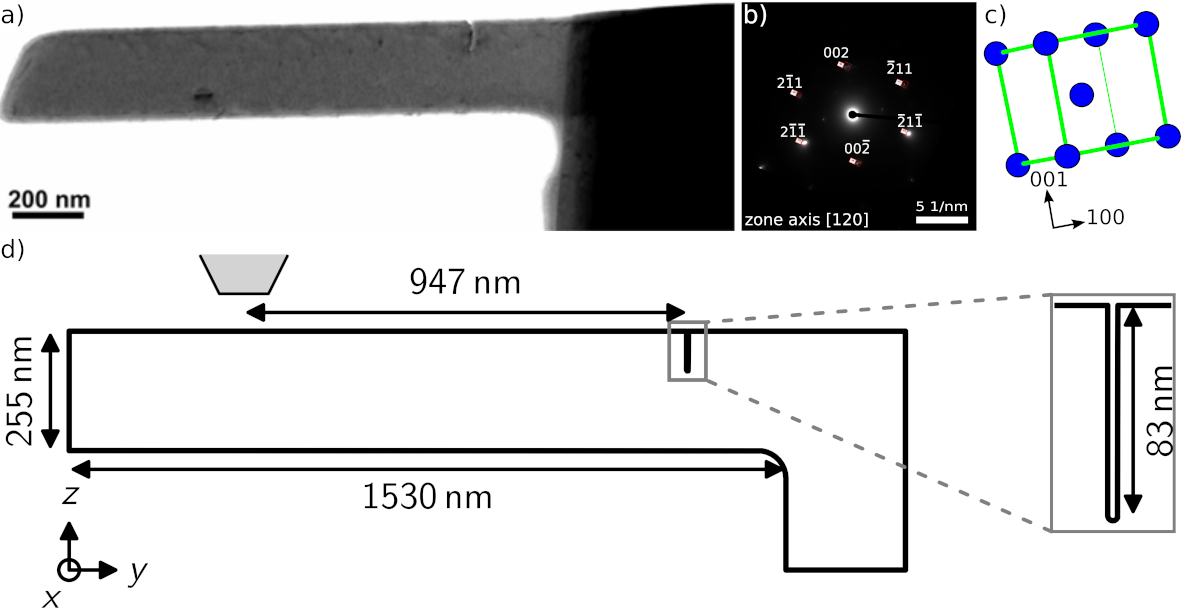}
	\caption{%
		a) STEM image of the nano-cantilever showing a nearly dislocation-free sample;
		b) Diffraction pattern of the sample showing the crystallographic orientation with zone axis \hkl[120];
		c) Scheme of the crystal orientation of the nano-cantilever;
		d) Sketch of the geometry of the notched micro-cantilever beam which is also used for the \glsfmtshort{finite element method} analysis. The thickness of the beam is \SI{187}{\nano\meter}; 
		the indenter is shown as  gray-shaded area; the radius of the notch is \SI{2}{\nano\meter}.
		The time-displacement and time-indenter-force curves obtained from the experiment are shown in  \cref{fig:time_displacement_force_curves}, selected frames from the recorded video are shown in \cref{fig:analyzed-frames}.
	}
	\label{fig:experiment-setup}     
\end{figure}

The \emph{in-situ} \ab{transmission electron microscopy} experiment  was performed on \ab{focused ion beam} milled single crystal Chromium (Cr) specimens. These samples
were prepared by initially grinding a Cr single crystal into a wedge shape. Using a \ab{focused ion beam} with \SI{30}{\kilo\electronvolt} gallium cations, a nano-beam geometry was fabricated,
\added{
	which has a cross-section with a height of \SI{255}{\nano\meter} and a thickness of \SI{187}{\nano\meter}.
}
The unavoidable surface damage of this preparation step was removed by annealing the specimens at \SI{900}{\celsius} for  \SI{20}{min} in the TEM. 
\oldadd{
	Throughout, temperature was monitored via an integrated thermocouple, and due to the slow heating thermal equilibration is presumed.
}
Finally, a notch with a radius of of about \SI{2}{\nano\meter} was introduced into the nano-beam via a condensed \SI{200}{\kilo\electronvolt} electron beam. \add{A STEM image of the sample and a sketch of the sample geometry is shown in \cref{fig:experiment-setup}a and \cref{fig:experiment-setup}d, respectively}. 
\delete{The thickness of the nano-beam is \mbox{\SI{187}{\nano\meter}}.}\ 
Before conducting the actual experiments, the crystallographic orientation was determined from diffraction analysis. The nano-beam is closely oriented along \hkl[100] and observed through \hkl[120] direction, as shown in \cref{fig:experiment-setup}(c).

\begin{figure}
	\centering
	\includegraphics[width=\columnwidth]{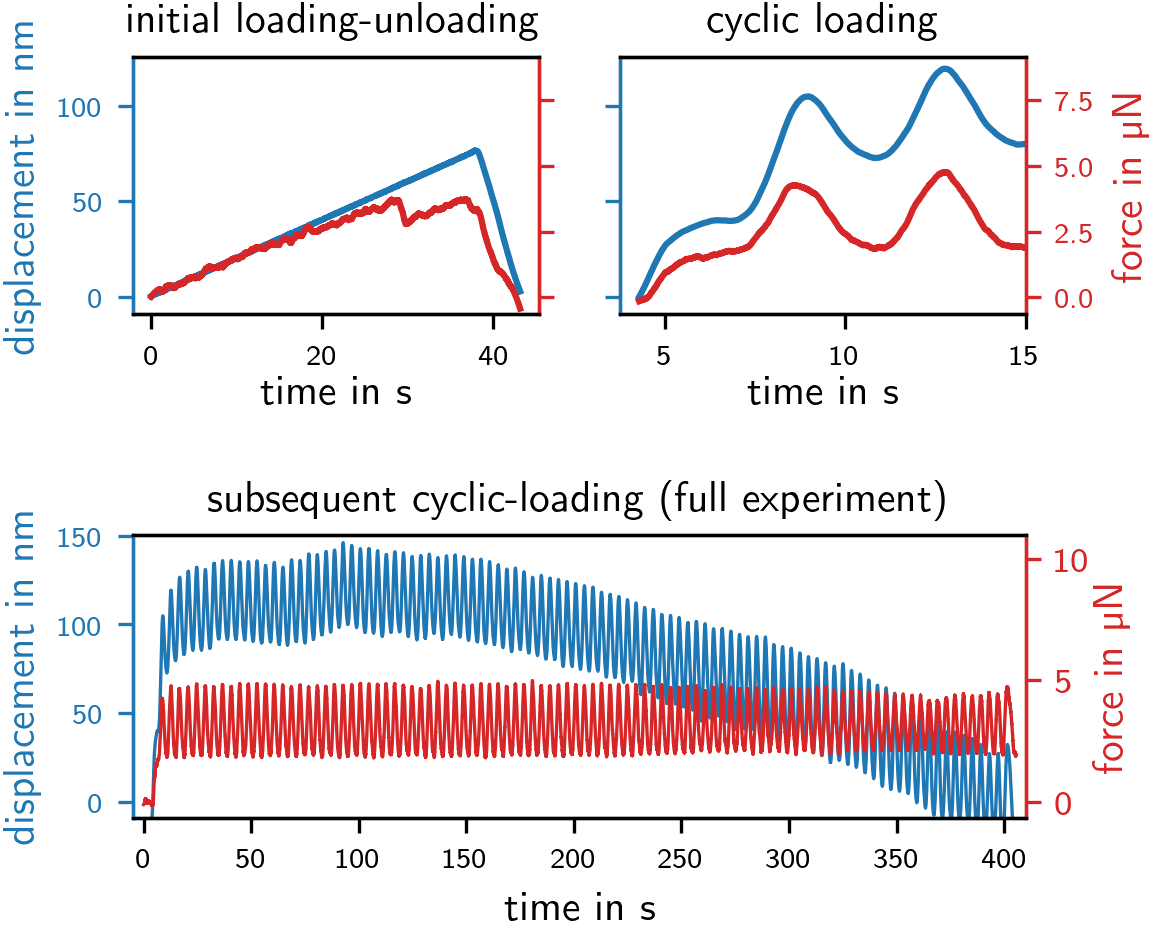}
	\caption{%
		Displacement (blue) and indenter force (orange) over time for the initial loading-unloading (left) and the subsequent cyclic-loading (right) experiments. The slope of the indenter displacement and the indenter force for the first ten seconds of the initial loading-unloading experiment are aligned to highlight deviations from linear behavior. 
	}
	\label{fig:time_displacement_force_curves}
\end{figure}

The indenter used for these in-situ experiments was a Hysitron Picoindenter \textsc{pi}-95 in a feedback-enabled displacement controlled mode.
Synchronized videos of the experiments were recorded in STEM mode with a frame rate of \SI{30}{\per\second} and a spatial resolution of \SI{1.28}{\nano\meter/pixel}.
Using this notched nano-cantilever, two experiments were performed in sequence.
During the initial quasi-static loading-unloading experiment, the indenter was displacement controlled at a constant rate of \SI{2}{\nano\meter\per\second}.
Upon first dislocation activity, the loading is stopped and the indenter is removed.
For the subsequent cyclic-loading experiment, the indenter was load controlled to cyclically load the nano-beam with a mean load equal to the load at which first dislocation activity was observed in the initial loading-unloading experiment, and an amplitude that is half of it with a frequency of \SI{0.25}{\per\second} and with  R-value (stress ratio between maximum and minimum load) of $R=0.5$. The corresponding displacement/load vs. time plots can be seen in \cref{fig:time_displacement_force_curves}.
More details on sample fabrication and characterization are provided in the Supplementary Information. 

\add{Identifying the dislocation character is not straightforward in this case.}\
\oldadd{
	While TEM is, in principal, ideally suited for this  using the well known $\vec{g}\times \vec{b}$ criterion, in the present case this is more challenging, as very thick samples are required. This limits the $\alpha$ tilting angles, and the second tilting direction is not available in the used holder.
	Here, the strength of the present data minig approach comes into play and allows to gain additional information from the in-situ images in the aftermath of the experiment.
}

\subsection{Reconstruction of the 3\textsc{D} dislocation microstructure}
\label{sec:analysis:slip-system-identification}
The next step consists in digitizing the TEM images such that dislocation lines are available as polygons.
To uniquely identify the dislocations, we assign them an identifier.
Those seen in frame \num{4788} of the initial loading-unloading experiment (cf. \cref{fig:analyzed-frames}) are assigned the numbers \numrange{1}{7} from right to left, i.e, in their order of nucleation from the notch surface.
Therefore, at frame \num{4915} of the same experiment we see the dislocations \numlist{4;3;2;1}, from left to right.
The additional dislocation seen in frame \num{308} of the subsequent cyclic-loading is the dislocation number \num{8}.

As we want to use the \ab{finite element method} to compute the stresses and strains in the three-dimensional configuration, we also require the three-dimensional configuration of the dislocations.
Thus, we need to reconstruct the three-dimensional geometry from the two-dimensional frames shown in \cref{fig:analyzed-frames}.
To this end, we first extract the two-dimensional configuration of each dislocation from the \ab{transmission electron microscopy} images in image coordinates.
Along with the dislocations, we also extract the scale bar, which allows us to convert the image coordinates to physical coordinates. In this work, the extractions of the digitized dislocation line coordinates are conducted with the open-source software \emph{labelme} \autocite{labelme_2016} which results in one polygonal chain per dislocation and per frame of the TEM movie.
Given the slip plane $\vec{n}$ of a slip system and a projection direction $\vec{d}$, we may project the $n_{v}$ vertices $\vec{v}_{i}$
of a polygonal chain onto the slip plane via
\begin{equation}
	\label{eq:polygonal-chain-projection}
	\vec{v}^{\prime}_{i}
	=
	\vec{v}_{i}
	+
	\frac{
		(
		\frac{1}{2}(\vec{v}_{1} + \vec{v}_{n_{v}}) - \vec{v}_{i}
		)
		\cdot
		\vec{n}
	}{
		\vec{d}
		\cdot
		\vec{n}
	}
	\;
	\vec{d}.
\end{equation}
The projection direction coincides with the viewing direction of the \ab{transmission electron microscopy} images.
Conceptually, \cref{eq:polygonal-chain-projection} yields the intersection point of a line and a plane.
The line is defined by \begin{enumerate*} \item the vertex $\vec{v}_{i}$ of the two-dimensional configuration, where the missing coordinate takes on an arbitrary value, and \item the viewing direction $\vec{d}$ \end{enumerate*}.
The plane is defined by \begin{enumerate*} \item the midpoint between the initial vertex $\vec{v}_{1}$ and the terminal vertex $\vec{v}_{n_{v}}$, and \item the normal vector \vec{n} of the slip plane \end{enumerate*}.
Their intersection point then is the vertex $\vec{v}^{\prime}_{i}$ of the three-dimensional dislocation configuration that corresponds to the vertex $\vec{v}_{i}$ of the two-dimensional one. A similar reconstruction has been done by one of the authors in \autocite{Oliveros2021_MatChemPhys272, Zhang2022}.

\subsection{Calculation of stresses of dislocation structures with the Finite Element Method.}
\Ab{finite element method} is a method for approximating and numerically solving partial differential equations and is particularly suitable for solid mechanics problems in finite domains, see e.g. \autocite{Fish2007}. Stresses and strains of dislocation microstructure can be computed with \ab{finite element method} due to the fact that the motion of a dislocation results in a plastically slipped area \autocite{Anderson_2017_a} that, together with a suitable physical regularization, can be translated into a contribution to an eigenstrain tensor field.  For this work the regularization of Jamond et al. \autocite{Cai2006_JMPS54} was used. \oldadd{More details on computing dislocation stress fields using the finite element method can be found in the Supplementary Information and}\ in \autocite{Cai2006_JMPS54, Po2018_IJP103, Jamond2016_IJP80}.The resulting elastic eigenstrain problem was solved by the \ab{finite element method}, based on a self-written code using the version 9.0 of the \emph{deal.ii} C++ Library \autocite{dealii2019design}. Mesh refinement was performed repeatedly with an Kelly error estimator \autocite{Kelly1983} ensuring that the results are mesh size independent and the quantities of interest converge. One advantage of this computational strategy is that the \ab{finite element method} automatically takes free surfaces into account and correctly represents image forces at surfaces and interfaces \autocite{Sandfeld2013_MSMSE21, SAMAEE2020138295}.
For the material model the cubic crystal symmetry of chromium and the resulting anisotropy in the stiffness tensor was taken into account for all \ab{finite element method} analyses.
The following elastic properties were used: $C_{11} =$\SI{339.8}{\giga\pascal}, $C_{12}=$\SI{58.6}{\giga\pascal}, and $C_{44}=$\SI{99.0}{\giga\pascal} \autocite{Sumer_1963_a}. Zero displacement boundary conditions were prescribed on the fare most surfaces  in positive $y$-direction and in the far most negative $z$-direction. To model the indenter, we prescribed traction boundary condition equivalent to an average force of \SI{3.69}{\micro\newton}.

\section{Results}

\subsection{Observations during the \emph{in-situ} TEM experiment}
\label{sec:experiment:results}

The \emph{in-situ} \ab{transmission electron microscopy} experiment was  performed on \ab{focused ion beam} milled single crystal Cr specimens in Bright Field Scanning TEM (STEM) mode. 
\delete{
	A sketch of the geometry is shown in Figure 1.
}\ 
Data was acquired with a tilt of a few degrees from the \hkl[120] zone axis and a diffraction vector $g=\hkl[002]$. The diffraction pattern in the \hkl[120] zone axis is shown in \cref{fig:experiment-setup}b. 
The observed dark dislocation lines with Burgers vectors $\vec{b}$ validate the visibility criterion $\vec{g}\cdot\vec{b} \neq 0$.

\begin{figure}
	\centering
	\includegraphics[width=\columnwidth]{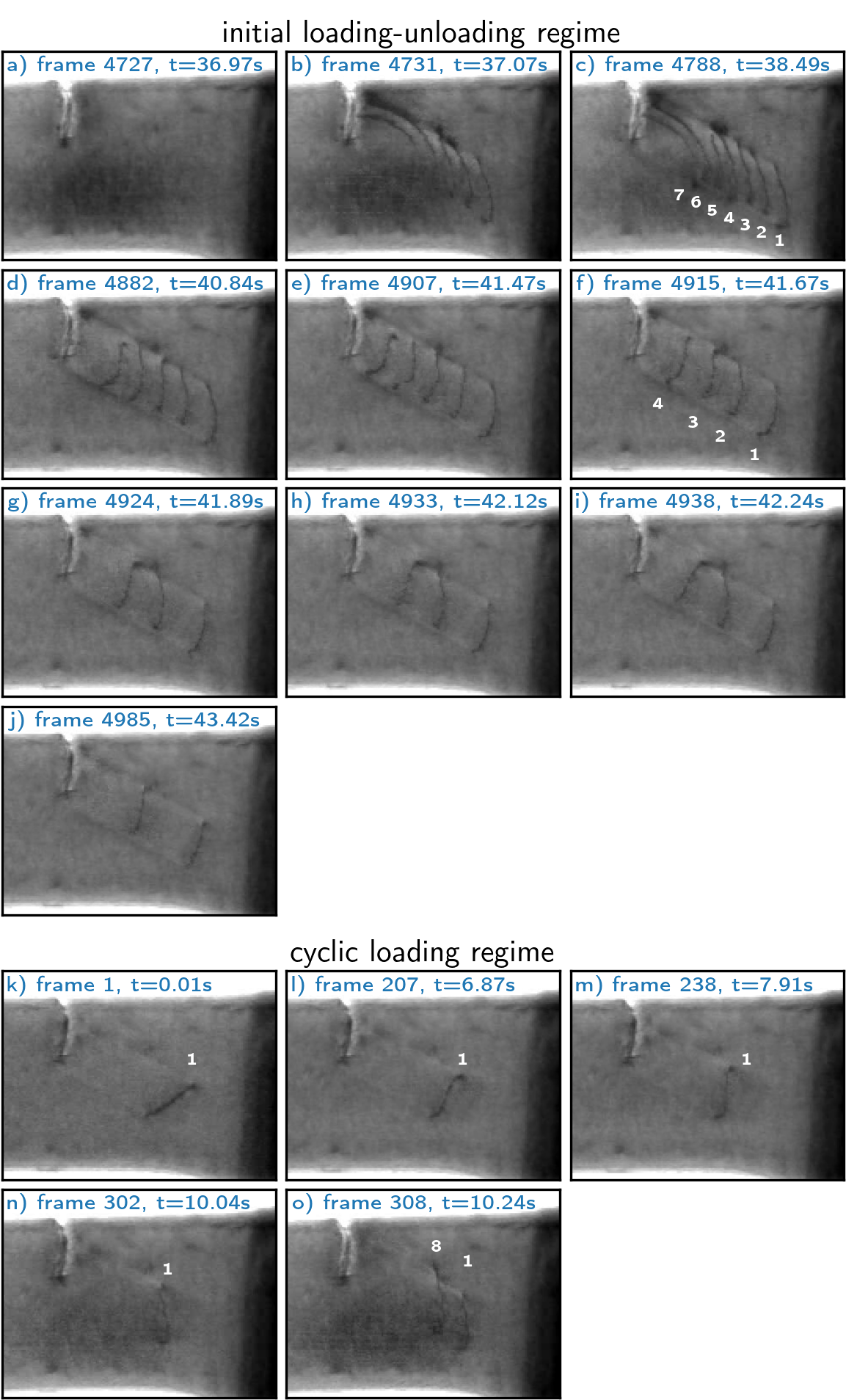}
	\caption{%
		\Ab{transmission electron microscopy} micrographs extracted from the \emph{in-situ} experiments showing the notch region of the specimen. The numbering of the dislocation always starts with the dislocation that got nucleated first. Frames a-j: quasistatic loading-unloading test. Frames k-o: cyclic loading.
	}
	\label{fig:analyzed-frames}
\end{figure}

The experiment consists of two different loading regimes: an initial quasi-static bi-linear loading-unloading regime and a subsequent cyclic loading sequence. Videos of the experiment during the two loading regimes are provided as Supplementary Online Material. 
The resulting time-displacement and time-indenter-force curves are shown in \cref{fig:time_displacement_force_curves}; selected frames from the video that was recorded during the experiment are shown in \cref{fig:analyzed-frames}.

During the initial loading-unloading regime in displacement-controlled mode, 
\replace{
	the latter increases linearly with time up to \mbox{\SI{76}{\nano\meter}} (and load \mbox{\SI{3.32}{\micro\newton}}) for \mbox{\SI{38}{\second}}; with a displacement rate of \mbox{\SI{2}{\nano\meter/\second}}.\
}{
	the displacement increases with a displacement rate of \SI{2}{\nano\meter/\second} for \SI{38}{\second} linearly up to \SI{76}{\nano\meter} and reaches a load level of \SI{3.32}{\micro\newton}. 
}\
This is followed by a linear displacement decrease down to zero over a time of \SI{5}{\second} (unloading). At \SI{29}{\second} after the beginning of the experiment, the force drops abruptly. In the recorded video (cf. Supplementary Online Material), 
we observe that the notch opens up at this point. Thereafter, the force continues to increase up to \SI{36.7}{\second}. At this point just before unloading, the first dislocation emission is observed. Consecutively, the force again drops sharply and after a brief force increase, the sample is unloaded.

In the subsequent cyclic-loading experiment, a load-controlled mode is used to cyclically load the nano-cantilever with a mean load equal to the load at which first dislocation activity was observed in the previous loading-unloading experiment, an amplitude half of this value, at a frequency of 
\SI{0.25}{\per\second}. The indenter comes into contact with the cantilever after $\approx$\SI{4}{\second}. Initially, the displacement increases up to \SI{40}{\nano\meter} until the lower limit indenter force of \SI{1.9}{\micro\newton} is reached. Notably, compared to the initial loading-unloading experiment, a larger displacement is required to achieve this force. The indenter is then attenuated to cyclically load the cantilever between \SIrange{1.9}{4.8}{\micro\newton} at \SI{0.25}{\per\second}. In the first few cycles, the required displacement to uphold the indenter force acting on the cantilever increases from an average value of \SI{89}{\nano\meter} in the \nth{1} cycle to \SI{114}{\nano\meter}  in the \nth{8} cycle at \SI{37}{\second}. After a brief plateau, the average displacement reaches a maximum of \SI{123}{\nano\meter} in the \nth{22} cycle at \SI{93}{\second}. Subsequently, the average displacement decreases until the end of the experiment.

The corresponding dislocation structures are shown in \cref{fig:analyzed-frames}\replace{ which show the extracted video frames that were analyzed in detail within this work. }{: the shown video frames are those analyzed in detail within this work.}\
Both in-situ tests were acquired and recorded in STEM mode with a frame rate of \SI{30}{\second^ {-1}}, thereby oversampling the mechanical experiment.

The micrographs are numbered alphabetically (a-j for the initial loading regime, k-o for the cyclic loading regime). Dislocations are assigned identifiers where $1$ is always the first dislocation that was nucleated in the first loading regime. Thereby, dislocations with the same numbers in different frames denote identical objects, as verified from the full video recorded. As the second loading regime (cycle) starts with a single remaining dislocation from the previous quasi static loading, this also remains as dislocation 1, and consequently the second dislocation seen in \frameO 
of the subsequent cyclic-loading will be denoted as number 8.

Within the initial loading-unloading experiment, \frameA shows the state of the micro-cantilever beam just before the first dislocation nucleates. 
After  consecutive nucleation of dislocations 2-7 in \frameC, unloading starts, the indenter load is decreased and all except one dislocation slip back towards the notch where they get absorbed by the notch surface. 
The final configuration of the quasi-static experiment is shown in \frameK 
of the subsequent cyclic-loading experiment. Upon applying the mean load again the dislocation moves slightly until a second dislocation nucleates from the notch just after \frameN. 

To identify the dislocations' slip system and their line directions a 
combined data mining strategy and FEM analysis is proposed in sections \ref{sec:analysis 1}~-~\ref{sec:analysis 3}. There, \enquote{classical} TEM analysis is additionally used for validation of the computational results in \cref{sec: analysis 2}. The computational strategy can be easily generalized to new situations and does not make or require any additional assumptions.
It provides detailed information of the three-dimensional dislocation microstructure  
\replaced{and is able to quantitatively describe the }{taking into account}  the actual internal stress state of the nano-cantilever using anisotropic elasticity and also considers possible geometrical imperfections of the experiment.

\subsection{Computational analysis 1: Geometrical consistency}
\label{sec:analysis 1}
Chromium is a \ab{body-centered cubic} material, thus 48 possible slip systems need to be considered. To narrow down the list of candidates, the argument in this first analysis step is a purely geometrical one: start and end points of the observed dislocations are located on two parallel lines, i.e.,the intersection lines of the slip plane with the front and back surface of the beam that can be seen in the \ab{transmission electron microscopy} images as slip traces. 

Applying the digitization and 3D reconstruction approach outlined in the \emph{Methods} section to all \ab{transmission electron microscopy} frames results in mathematical polygons in three-dimensional space that represent the dislocations \emph{for each of the slip systems}. The projection of this 3D dislocation polygon into the $x-y$--plane might or might not fit into the thickness of the beam.

\begin{figure}
	\centering
	\includegraphics[width=\columnwidth]
		{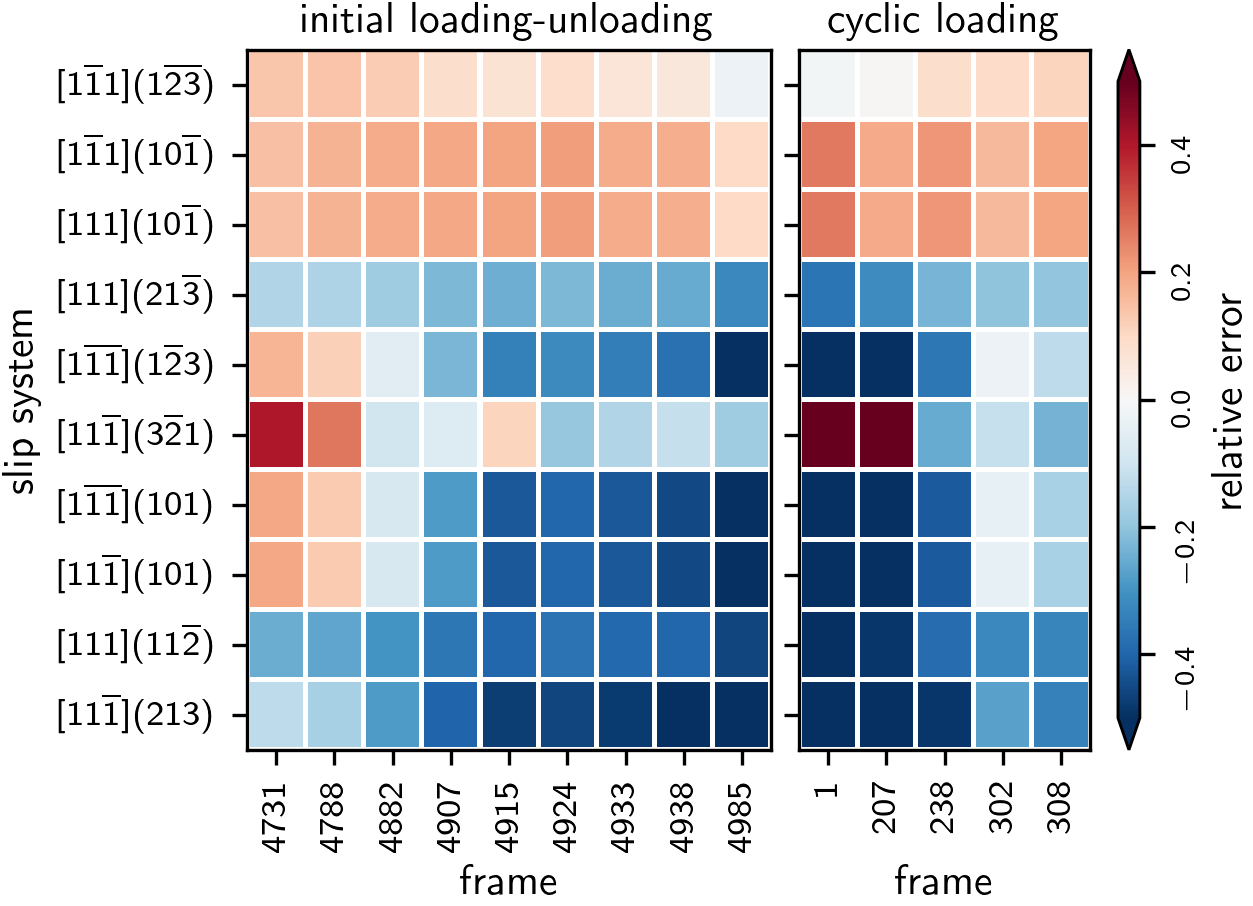}
	\caption{
		Consistency of the projection of dislocation \num{1} with the beam thickness for frames b-o:
		depending on the slip system candidate, the 3D reconstructed line would not exactly terminate at the beam surface. The shown relative error is this deviation normalized by the beam thickness of \SI{187}{\nano\meter}. A plot for all possible slip systems can be found in the SI.
	}
	\label{fig:projected_dislocation_depth_1}
\end{figure}

Aligning the start point of the dislocation polygon with the front of the nano-cantilever results in a more or less significant gap between dislocation end point and the back of the beam, which changes depending on the 
slip system. This deviation, normalized by the beam thickness, is used as an error measure making slip systems with smaller relative error better candidates \add{to be the slip system on which the dislocations are located in reality}. An additional error arises from manually digitizing the dislocations from the images, which is why it is very beneficial being able to analyze numerous data sets from multiple frames.
Analyzing the deviations for dislocation \num{1} in all but the first, dislocation-free \frameA is shown in \cref{fig:projected_dislocation_depth_1} for selected slip systems, while a complete plot encompassing all slip systems can be found in the Supplementary Information. 
Here, the slip systems are sorted with respect to their average error over all frames. Most slip systems can be excluded due to a fairly large error (e.g., a value of $0.3$ implies that the reconstructed dislocation ``sticks out'' of the beam by roughly a third of its thickness).
Furthermore, a number of slip systems not shown in the figures have slip planes aligned parallel to the viewing direction, so they can also be excluded.

Subsequently, only the five slip systems with the lowest average error, i.e., the top five rows of \cref{fig:projected_dislocation_depth_1}, are considered. Thus, the best candidate slip systems for the dislocation number \num{1} are:
\begin{itemize}
	\vspace{-0.3em}
	\setlength\itemsep{0em}
	\item \hkl[1-11]\hkl(1-2-3) and \hkl[1-11]\hkl(10-1)
	\item \hkl[111]\hkl(10-1) and \hkl[111](21-3)
	\item \hkl[1-1-1]\hkl(1-23)
	\vspace{-0.3em}
\end{itemize}

\subsection{Computational analysis  2: Resolved shear stress just before initial dislocation nucleation}
\label{sec: analysis 2}

\begin{figure}
	\centering
	\includegraphics[width=\columnwidth]
	{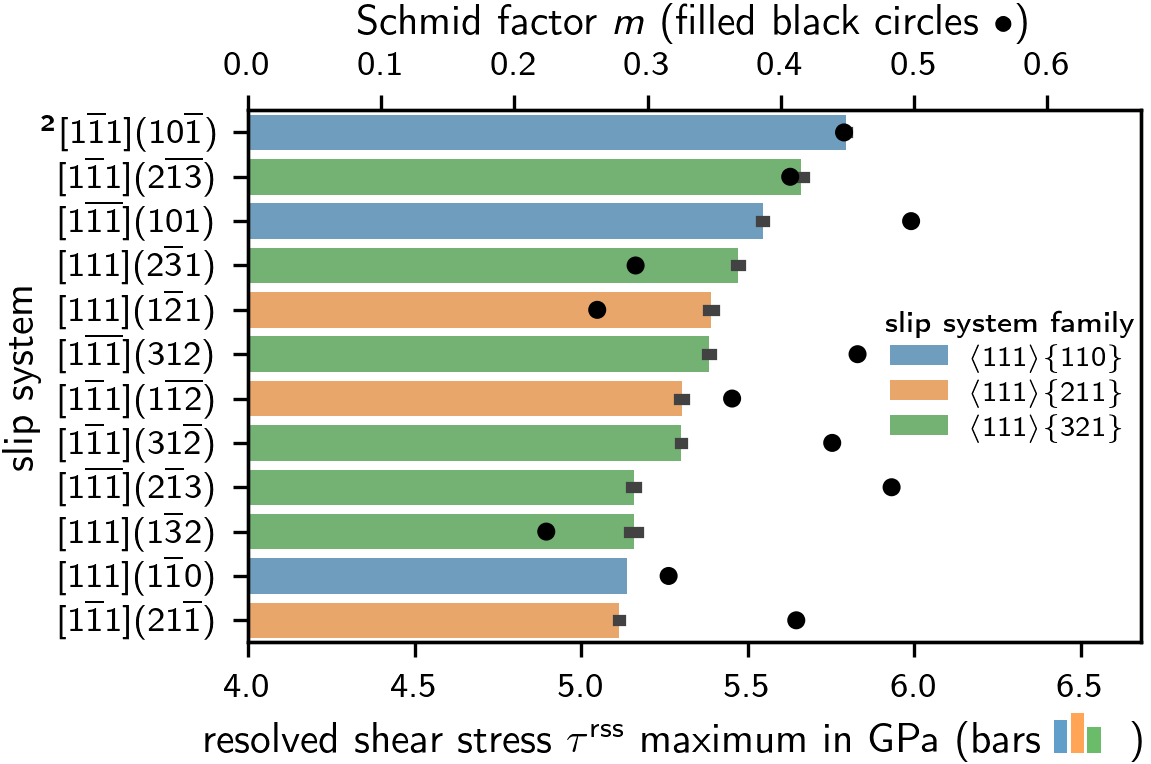}
	\caption{%
		Schmid factor (bullets) and resolved shear stress maximum (colored bars) at the notch surface for chosen slip systems observed in \frameA 
		during the initial loading-unloading experiment, just before the first dislocation nucleation.
		Bar colors correspond to the slip system family.
		Error bars denote the standard deviation of the resolved shear stress maximum for different misalignment errors.
	}
	\label{fig:resolved_shear_stress_1}
\end{figure}

To further narrow down the list of potential slip systems, an analysis of the stress fields for the situation of \frameA 
is performed. This is during the initial loading-unloading experiment just before dislocation \num{1} was nucleated at the notch surface. During dislocation nucleation, a high resolved shear stress that acts on the dislocation as driving force, away from the notch, is required. Therefore, by comparing the resolved shear stress acting on all possible slip systems one can exclude those with comparably low values as potential slip system candidates.

A commonly used method of estimating resolved shear stresses is an analysis using Schmid factors $0\leq m\leq 1$ (the higher the value $m$ the higher the resolved shear stress) for axial loading of a specimen. The bending deformation of the nano-cantilever results in large tensile stresses in the $y$-direction that are significantly larger than the other stress components such that this situation can be locally, near the notch tip, treated as axial tension. \Cref{fig:resolved_shear_stress_1} shows Schmid factors for some exemplary slip systems as black bullets (a complete plot with data for all slip systems can be found in the Supplementary Material).

None of the previously identified slip system candidates is among those with the highest Schmid factors: e.g. the Schmid factor of the \hkl[1-11]\hkl(10-1) system is more than 10 \% below that of the \hkl[1-1-1]\hkl(101) system having the highest Schmid factor. 
To understand this apparent contradiction and to analyze if the triaxiality of the stress state needs to be considered, a full \ab{finite element method} analysis with an anisotropic elasticity material model was performed. Automated mesh refinement around the notch was used to ensure that the stress fields are completely resolved. The bottom of \cref{fig:experiment-setup} shows the idealized geometry of the simulation model. Further details of the \ab{finite element method} simulation are given in the Methods section above and in the Supplementary Information.

\replaced{Based on such \ab{finite element method} calculations}{With this setup}, the resolved shear stress can be computed -- just before the nucleation of the first dislocation in the initial loading-unloading experiment --  for all slip systems. For each of those, the  stress maximum that is located near or at the notch surface was identified. The 12 \ab{body-centered cubic} slip systems that exhibit the highest resolved shear stress maxima are shown in \cref{fig:resolved_shear_stress_1} as colored bars.
The superscript in front of the topmost slip system \replaced{indicates that this is the second}{refers to the second} of the five slip systems identified as the best candidates before, compare \cref{fig:projected_dislocation_depth_1}: the slip system \crystaldirection{1 -1 1}\crystalplane{1 0 -1} exhibits the highest observed resolved shear stress maximum with \SI{5.79}{\giga\pascal}.
This is \SI{2.3}{\percent} higher than the next highest observed resolved shear stress maximum overall, and \SI{22}{\percent} higher than for the next highest slip system candidate for this dislocation, \crystaldirection{1 -1 1}\crystalplane{1 -2 -3} (an extended plot with data of all \ab{body-centered cubic} slip systems can be found in the Supplementary Material).
While the Schmid factors of all possible slip systems very roughly follow the trend of the resolved shear stresses, the fluctuations between them are significant. This indicates that the underlying uniaxiality assumption for using the Schmid factors is strongly violated; only a computational model that takes the full triaxial stress state at the notch tip into account, should be used for a detailed analysis.

\begin{figure}
	\centering
	\includegraphics[width=0.83\columnwidth]{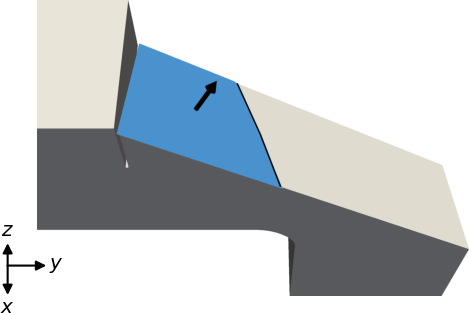}
	\caption{%
		Spatial location of the first dislocation (shown as thin black line) and the plastically slipped  area (shown as blue area). The thick black arrow indicates the direction of the slip plane normal. The material above the inclined slip plane, on the right side of the notch was removed for visualization purposes.
	}
	\label{fig:dislocation-sense-definition}
\end{figure}

Furthermore, the actual sample does not have an atomically flat surface, the indenter might not be perfectly perpendicular to the surface, or the measurement of the crystallographic orientation of the crystal comes with an error. To demonstrate how such experimental imperfections influence the resolved shear stresses, the small, black bars at the end of the color bars in \cref{fig:resolved_shear_stress_1} represent error bars due to alignment errors. These result from computations with slightly misaligned crystallographic orientations: the experimentally measured angle is \ang{8}; it is the angle between the loading direction and the \hkl[001] axis of the crystal.
Therefore, seven additional simulations were conducted with values in between \ang{5} to \ang{11}. Compared to the difference among slip systems, the possible error resulting from the experimental uncertainty is very small. Thus, the experimental results tend not to be very sensitive with respect to small deviations of geometrical details of the sample or indenter which also justifies the inherent simplifications of such a simulation model.

Following the above analysis of the resolved shear stress maxima, it can be conclude that
the most likely slip system for the nucleation of the first dislocation is \crystaldirection{1 -1 1}\crystalplane{1 0 -1}. With this information it is now possible to fully reconstruct the geometry of the slip plane and the 3D dislocations as visualized in \mbox{\Cref{fig:dislocation-sense-definition}} for the first frame of the cyclic loading regime.
	Can this be validated by a \enquote{classical} TEM analysis? As such a manual analysis is laborious, a few simplifiying assumptions need to be made: (i) we only consider the \hkl<111>\hkl{110} system as it is the  slip system on which most commonly dislocation acitivity in BCC crystals takes place; (ii) taking into account the symmetries in cubic systems, only 6 out of 12 possible slip systems have to be crystallographically analyzed. This was done with a dedicated crystallography visualization software (see Supplementary Fig. 3), which (iii) has to make the assumption that dislocations are either purely edge or purely screw dislocations. One of the resulting most likely slip systems is the
	\textonehalf\hkl[1-11]\hkl(10-1) slip system. Further details are given in the Supplementary Material.

\subsection{Computational analysis 3: Peach-Koehler force}
\label{sec:analysis 3}

\begin{figure}
	\centering
	\includegraphics[width=\columnwidth]{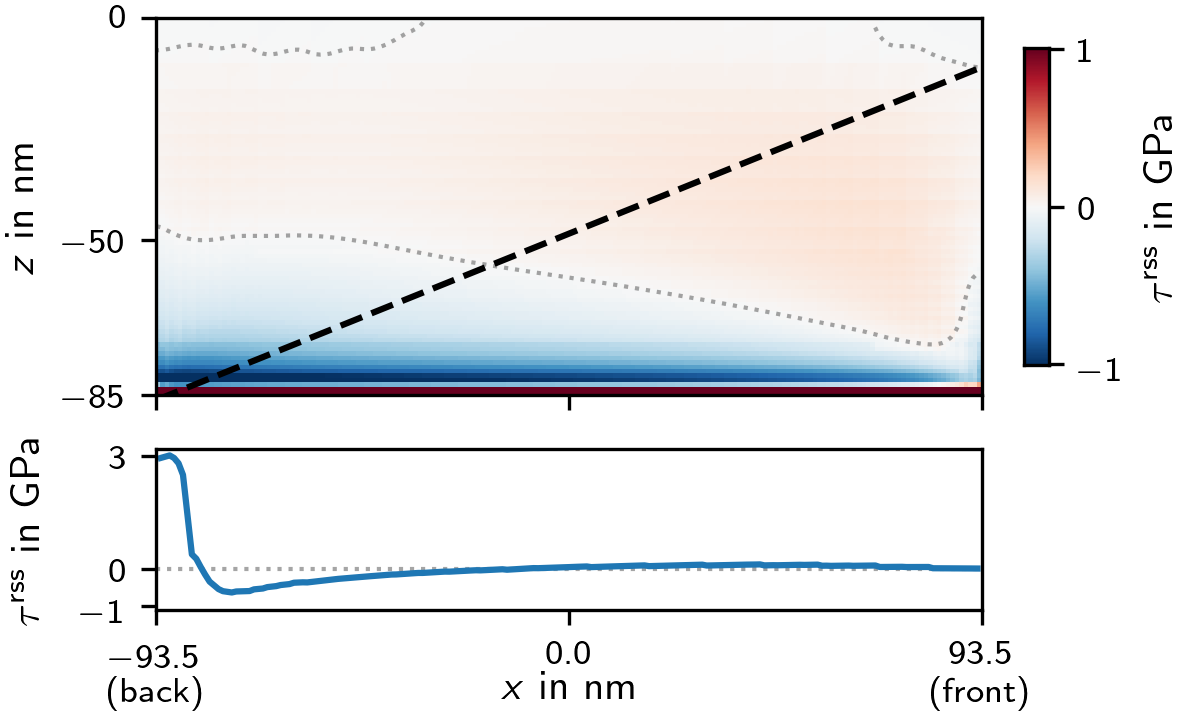}
	\caption{
		Resolved shear stress $\tau^{\mathrm{rss}}$ of the \crystaldirection{{1 -1 1}}\crystalplane{{1 0 -1}} slip system  at the right notch surface (i.e., the plane with normal pointing into positive $y$ direction) (top plot). The dashed line indicates the intersection of the notch with the slip plane (cf. \cref{fig:dislocation-sense-definition}). The bottom plot shows the resolved shear stress along this intersection line. 
		Thin dotted lines denote zero values of the stress. Sign conventions are such that a positive $\tau^{\mathrm{rss}}$ acts to drive a dislocation with a right-handed line sense away from the notch towards the right.
	}
	\label{fig:notch-slip-system-peach-koehler-force-1} 
\end{figure}

A dislocation is characterized by its Burgers vector $\vec{b}$ and a line direction vector $\vec{\xi}$ that varies along a curved line. Knowing the \replaced{exact crystallographic details of the slip system/slip plane automatically determines the Burgers vector as well. This together with }{slip system/slip plane and}   the three-dimensional reconstruction of the points of the line from the \ab{transmission electron microscopy} micrographs is not sufficient to fully recover the mathematical dislocation representation as this \replaced{does not directly yield the}{only yields the line tangent but not the} line direction vector. Thus, it leaves the line \emph{sense} undetermined. The line sense, however,  is an important piece of information for details of the plastic deformation because dislocations with opposite line sense and otherwise identical properties would move into opposite directions.

Throughout this determination of the line sense we use the convention that if the thumb of the right hand points along the slip plane normal $\vec{n}$ then the other fingers curling  around $\vec{n}$ indicate the direction of a ``right-handed line sense''. Otherwise the dislocation has a ``left-handed line sense''. Closely related is the \ab{Peach-Koehler force}, $\vec{f}=(\vec{b}\cdot\vec{\sigma})\times \vec{\xi}/||\vec{\xi}||$ with $\vec{\sigma}$ the resolved shear stress tensor: reversing the line sense (i.e., rotating it by \ang{180} in the plane) results in a change of sign of $\vec{f}$, and thus, the \ab{Peach-Koehler force} can be indirectly used to determine the line sense.

Subsequently, the \ab{finite element method} model is used to obtain the direction and value of the \ab{Peach-Koehler force} that would act on the dislocation on the respective slip plane close to the notch. The resolved shear stress that the first dislocation would experience at the surface of the notch in \frameA is shown in the bottom figure of \cref{fig:notch-slip-system-peach-koehler-force-1}. 
The highest value of the resolved shear stress of \SI{+3.01}{\giga\pascal} is observed at the notch tip (which is in the figure at $x\approx$\SI{-93}{\nano\meter}).
With increasing $x$ along the intersection, the value decreases down to the minimum value of \SI{-0.46}{\giga\pascal} before it then tends towards \SI{0}{\giga\pascal}.
Given the positive sign of the peak stress a dislocation with a right-handed line sense would experience a force that drives it away from the notch and further into the material. A left-handed dislocation line sense would result in the opposite direction of motion. 

Therefore, the results from these three computational analysis steps can be summarized as follows:
the dislocation \num{1} is located on a\crystaldirection{{1 -1 1}}\crystalplane{{1 0 -1}} slip system; it is a mixed dislocation and has a right-handed line sense, i.e., the line direction vector of the dislocation in \cref{fig:notch-slip-system-peach-koehler-force-1} would point from the front towards the back of the sample.

For validation purposes and as a generalization the analysis was also applied to frames with more dislocations: the analysis for dislocation 8 can be mostly done in analogy to that of dislocation 1 and yields the same result. To resolve an occurring ambiguity, the geometry slip traces on the surfaces of the specimen were additionally considered.  Analyzing  dislocations \numrange{2}{5} was, due to limited \ab{transmission electron microscopy} data, only partially possible but still supported the choice of the already identified slip system. More details can be found in the Supplementary Material, which also contains an analysis of an idealized  version of the dislocation pile-up shown in \cref{fig:analyzed-frames}.

\section{Discussion}

\subsection*{Dislocation nucleation and their evolution}
\label{sec:analysis:nucleation-of-dislocations}
\begin{figure*}
    \centering
    \includegraphics[width=0.8\textwidth]{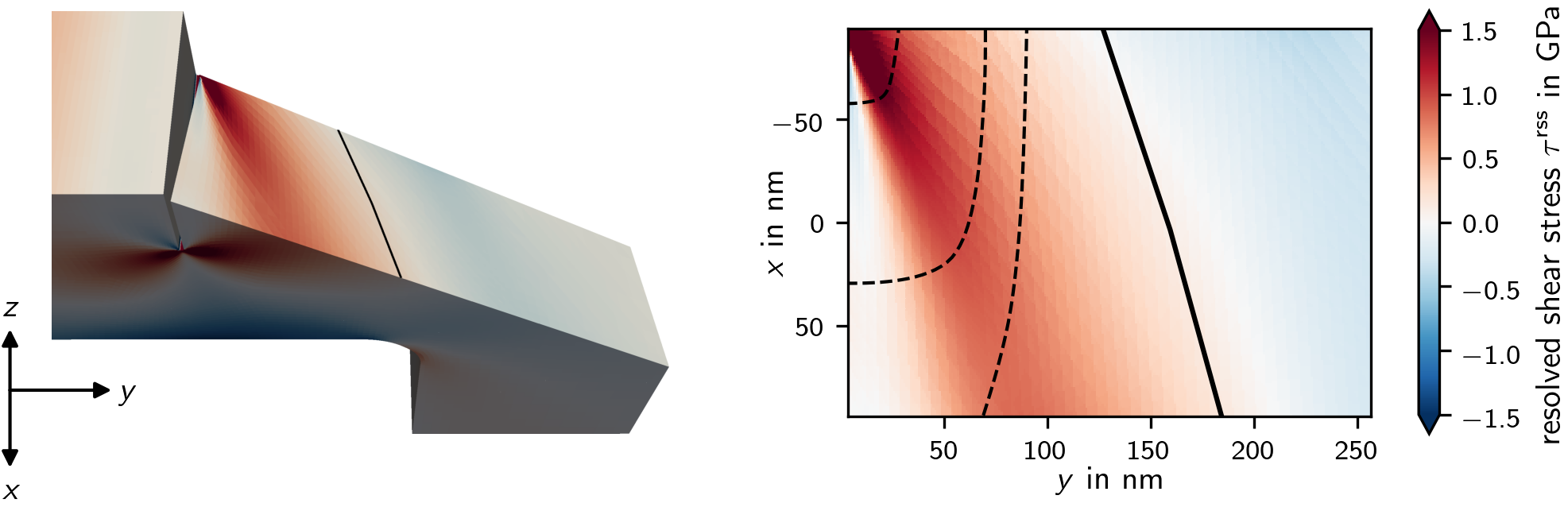}
    \caption{%
        Resolved shear stress for the \crystaldirection{{1 -1 1}}\crystalplane{{1 0 -1}} slip system. 
        The stress results only from the indenter loading, the dislocation stress is not superimposed.
        Left: Cut through the sample along the slip plane; right: stress in the slip plane, viewed along the $z$-axis.
        The dislocation  \num{1} (black solid line)
        of the cyclic-loading experiment is shown as black, solid line in both plots. Dashed lines sketch some possible intermediate configurations of the dislocation after the nucleation.
    }
    \label{fig:dislocation-1-slip-plane-rss}
\end{figure*}

Based on the results and analysis so far, one may now attempt to understand how the first dislocations are nucleated from the notch tip and how they move afterwards.
As seen in \cref{fig:notch-slip-system-peach-koehler-force-1} (and the supplementary Fig. 8), 
the dislocations only exhibit forces pushing them into the material at the tip of the notch.
Slightly above the notch tip, dislocations are then experiencing a force that pulls them towards the surface.
For a better idea of the spatial details of these driving forces, the resolved shear stress on the slip plane of dislocation \num{1} is shown in \cref{fig:dislocation-1-slip-plane-rss}.
There, a positive resolved shear stress acts to expand this dislocation, driving it into the material whereas a negative shear stress drives it towards the notch. 

The only place where a just nucleated dislocation is able to expand is therefore at the tip of the notch (in the right panel of \cref{fig:dislocation-1-slip-plane-rss} the stress maximum at the upper left corner).
During the initial expansion, the small segment of the dislocation that ends at the notch surface experiences small negative stresses that act to push it back towards the notch tip, thereby anchoring it while the rest of the dislocation expands.
Acting against this is the line tension, a thermodynamic force that acts to reduce dislocation line length.
The part of the dislocation that is oriented roughly in $x$ direction is being pushed into the material and at the same time \enquote{drags along} the part anchored at the notch surface.
Eventually, this leading part reaches the front side of the beam and rotates -- and leaves a slip trace on the surface. 
Further into the material, the resolved shear stress decreases and at about $y \approx$\SI{150}{\nano\meter} changes sign.
This is similar to the neutral axis of a regular bending beam.
The configuration of the dislocation \num{1} for \frameK of the cyclic-loading experiment is shown as black line. It aligns well with the neutral axis of this beam. This also explains why the dislocation did not get pulled back towards and absorbed by the notch surface unlike the others when the load was decreased in the initial loading-unloading experiment: the dislocation is in a metastable position.
A possible sequence of positions of this dislocation is shown as dashed lines in \cref{fig:dislocation-1-slip-plane-rss} (this was obtained by interpolation considering the stress field and not by a ``dislocation dynamics simulation''). The reconstructed motion of the dislocation also agrees well with the analysis of the stress intensity of the pileup presented in the Supplementary Information. 

\subsection{Can the nucleation process be understood by energy arguments alone?}
Now that slip system, line orientation and line sense are determined the energy-related properties of this system can be assessed in more detail. E.g., one can determine whether the nucleation of a dislocation is energetically favorable and results in global minimization of the elastic strain energy. 

\begin{figure}
	\centering
	\includegraphics[width=\columnwidth]{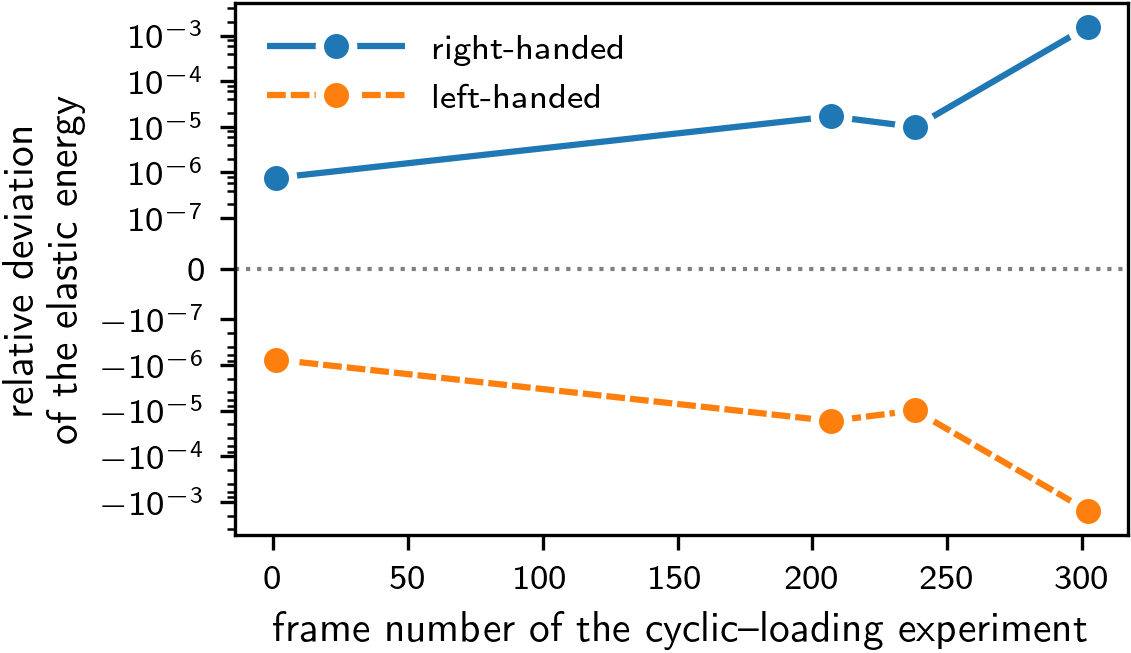}
	\caption{%
		Relative deviation of the total elastic energy for frames number 1, 207, 238, and 302, including only dislocation 1. The two lines show results for the two possible dislocation line senses.
	}
	\label{fig:beam_total_energies}
\end{figure}

The \ab{finite element method} model is again used to compute the total elastic energies for those static and cyclic frames in which only the first dislocation is present, i.e., frames \num{1}, \num{207}, \num{238} and \num{302}. Since we want to compare data from all these configurations we introduce a normalized measure for the energy deviation:
There, the reference value $\langle E^{\mathrm{el}}\rangle$ is the average total energy of the elastically deformed systems from all four considered frames without accounting for the dislocations, i.e., the state that only results from the indenter load (which varies for each frame). 
Then, the normalized deviations from this reference value are, on average, given by
$(E^{\mathrm{tot}}(\vec{\xi}) - \langle E^{\mathrm{el}} \rangle) / \langle E^{\mathrm{el}} \rangle$, where $\vec{\xi}$ is the line orientation and $E^{\mathrm{tot}}(\vec{\xi})$ is the spatially averaged energy for a particular frame resulting from indenter force and dislocation. The results are shown in \cref{fig:beam_total_energies}.
The solid blue line denotes the relative energy change due to the introduction of a dislocation with the above identified, positive line sense. The positive and increasing value might seem surprising as this suggests that also the total energy of the system is increased -- as opposed to the data for the dislocations with a left-handed sense (dashed line). However, the dislocations in these frames are only stationary for particular externally applied stress states and move back towards the notch, driven by image forces from the free surface and back-stresses among themselves, until they get absorbed at the notch once the indenter is released (cf. the supplementary video). Therefore, if the cyclic experiment is seen as a whole, this plastic deformation state is rather an \emph{anelastic} state, where the plastic deformation is not permanent but rather reversible.

%
%
\subsection{Summary and conclusion}

In this work, an in-situ TEM nano-fracture experiment was performed to detail the early dislocation nucleation events taking place from a sharp notch tip in a Cr single crystal. It was evidenced that toughening took place by nucleation and pile-up of dislocations from the crack tip. Upon unloading, all except one dislocation situated at the neutral axis annihilated at the crack surface they emitted from. During cyclic loading, a highly repeatable anelastic dislocation behavior in front of the crack tip was observed. 

To analyze the dislocation structure, their interaction and evolution in details, \olddelete{a computational analysis and data-mining approach was developed that is able to fully reconstruct each dislocation in 3D from in-situ micrographs.}\oldadd{a computational approach was developed that is able to (i) fully reconstruct each dislocation in 3D from in-situ micrographs and (ii) to compute the stress and strain energy density state in the specimen due to the presence of dislocations, the external loading and boundary conditions. Additionally, a data-mining framework was developed that helped to systematically sample the parameter space of possible slip systems and line orientations.}\ The exact identification of the active slip system and the line sense of dislocations became possible through a chain of physical arguments that were incorporated into the data analysis. A FEM-based stress analysis using the 3D dislocation structure and considering effects from image forces at the crack surface revealed details of the resolved shear stress, the driving force acting on the dislocations and the internal energy, while at the same time being rather insensitive to unavoidable minor experimental imperfections. Thus, this concept allows for active slip system and dislocation identification in situations of complex stress states and for massive amounts of data as encountered e.g. during in-situ experiments.

\oldadd{
	The tight combination of advanced microscopy techniques with computational analysis approaches is a very attractive combination and, in the future, will become more and more able to extract otherwise inaccessible information from in-situ experiments. Furthermore, systematic data mining with tailored analysis methods will help to create statistically relevant data sets, and thereby allows to identify patterns, trends and correlations, and helps to remove any human bias while preserving the important domain knowledge.}\ First steps into such a direction can be seen, e.g., in \cite{Zhang2022}.
Additionally, such reconstructed 3D dislocation microstructures could, in the future, be used as input or for direct validation of, e.g., discrete dislocation dynamics simulations or even of molecular dynamics simulations. Automating all analysis steps, e.g., through Deep Learning-based binary segmentation of microscopy micrographs \cite{Bulgarevich_2018_a, Trampert20211_Crystals11} will then also allow for high-throughput data analysis pipelines of microscopy data and will thereby even couple simulation and microscopy tightly.

\section{Authors' Contributions}
D.S.: software, investigation, formal analysis, validation, writing - original draft, visualization;
I.I.: investigation, writing - review \& editing;
R.S.: writing - review \& editing, validation;
P.I.: investigation;
S.S.: conceptualization, writing - original draft, supervision, visualization, funding acquisition, data curation;
D.K.: conceptualization of the experiment, supervision, writing - review \& editing, funding acquisition

\section{Competing Interests}
The authors declare no competing financial or non-financial interests.

\section{Data and Code Availability}
The data of all shown plots together with the code for reproducing the plots as well as supplementary data sets are available at \url{https://gitlab.com/computational-materials-science/xzy}. An additional snapshot of the repository with data and code has been published at the permanent doi \url{https://zenodo.org/xyz}. 

\section{Acknowledgments}
D.S. and S. S. acknowledge financial support from the European Research Council through the ERC Grant Agreement No. 759419 (MuDiLingo: 'A Multiscale Dislocation Language for Data-Driven Materials Science'). I.I. and D.K. acknowledge funding by the European Research Council under ERC Grant Agreement No. 771146 (TOUGHIT:  'Tough Interface Tailored Nanostructured Metals').

%
%
\printbibliography

\end{document}